\documentclass[prb,showpacs,floatfix,twocolumn]{revtex4}
\usepackage{graphicx}
\usepackage{amssymb}
\usepackage{dcolumn}
\usepackage{bm}
\begin{document}

\title{A forest-fire analogy to explain the {\em b}-value of the Gutenberg-Richter law for earthquakes}
\author{E. A. Jagla}

\affiliation{Centro At\'omico Bariloche and Instituto Balseiro, Comisi\'on Nacional de Energ\'{\i}a At\'omica, 
(8400) Bariloche, Argentina}

\begin{abstract}

The Dr\"ossel-Schwabl model of forest fires can be interpreted in a coarse grained sense as a model for the stress distribution in a single planar
fault. Fires in the model are then translated to earthquakes. I show that when a second class of trees that propagate fire only after some finite time is introduced in the model, secondary fires (analogous to aftershocks) are generated, and the statistics of events becomes quantitatively compatible with the Gutenberg Richter law for earthquakes, with a realistic value of the $b$ exponent. The change in exponent is analytically demonstrated in a simplified percolation scenario. Experimental consequences of the proposed mechanism are indicated.

\end{abstract}
\maketitle

Many physical systems react upon a continuous input of energy, by releasing the accumulated energy in discontinuous bursts, that are called avalanches in general, occurring when some threshold condition is reached. Examples range from sand piles, to magnetic domain inversions in ferromagnets, stress release on the earth crust in the form of earthquakes, and many others. 
Many times these  avalanches display a broad size distribution, typically following a power law, that is taken as a manifestation of the lack of intrinsic spatial scale in the system. Although in real systems this fact can only be approximate, it is usually taken as the ideal target for theoretical models that try to mimic the observed behavior. This concept has gone under the name of self-organized criticality\cite{soc}.
However, in order to describe a natural phenomenon that displays power law distributions, it is not necessary for theoretical models to be self organized critical. To be considered realistic, they need only to display a phenomenology consistent with observation, which is always limited, in spatial and temporal scales. 

For the case of earthquakes (EQs), the experimental evidence indicates that they follow the so called Gutenberg-Richter (GR) law\cite{gr,scholz}, stating that the number of EQs in a given magnitude interval $dM$ follows the empirical relation $N(M)\sim 10^{-bM}$.
This has been observed to be accurate in a large range of magnitudes with values of $b$ ranging from 0.8 to 1.2. When expressed in terms of the seismic moment $S$ (such that, up to a constant, $M=\frac 2 3\log_{10} S$), this law can be recast in the form $N(S) \sim S^{-\tau}$ with $\tau=1+\frac 23 b$, so realistic values of $\tau$ are in the range 1.5-1.8. Statistical models devised to describe EQs\cite{kawamura}
(most remarkably, the Burridge-Knopoff \cite{bk}, and Olami-Feder-Christensen\cite{ofc} models) typically consider the case of a single planar fault. GR law has been found in this kind models, and even a correct value of $\tau$ has been obtained in particular cases, typically by adjusting some parameters of the model, what goes against the robustness observed experimentally in very different conditions. 

In the last years it was realized that even if a correct GR law is reproduced, the sequences of events generated by the models mentioned above are unnatural in the following sense. Real EQ sequences display the peculiar feature of aftershocks (ASs), namely events that appear exclusively as a consequence of previous large events, and that are not directly correlated to the driving in the system. The existence of ASs indicates that there is some other important time scale at play, in addition to the one imposed by the external driving. The physical mechanisms associated to this internal time scale is related to plastic effects, creep dynamics, etc., and I refer to them in general as relaxation mechanisms.

Models including this kind of relaxation have recently been introduced\cite{hainzl,jagla_jgr,jagla_pre}. It was shown that they are able to produce ASs with realistic features. Most remarkably, when ASs occur, the exponent of the GR law was observed to change, and set to a value close to $\tau\simeq 1.7$, without fitting any parameter. The reason of this fact has not been adequately discussed up to now.

An appropriate scenario for this discussion turns out to be the Dr\"ossel-Schwabl model of forest fires\cite{ds}. Although introduced to model a very different problem, it is recognized that a coarse grained version of the model may qualitatively represent the evolution of the stress in a single planar fault system.\cite{pacheco,jensen,turcotte0} I summarize here the rules of the DS model for completeness. In an initially empty, two dimensional square lattice, sites are occupied randomly by trees, one every time unit $t_0$. After $r$ tree insertions, a lightning event occurs at some random position, burning the tree that is located there, and all trees that can be reached from this site through nearest neighbors occupied sites. This instantaneous burning defines the fires in the system, their size (the number of burnt trees) being denoted by $S$. In the limit in which $r\to \infty$ there are fires with arbitrarily large values of $S$. There is evidence that the system does not become truly critical in this limit\cite{grassberger,jensen2}, however a size distribution of fires with an exponent  around $\tau\sim 1.2$ has been consistently observed for a wide range of sizes, and this is sufficient for the  analysis presented here. 

A coarse grained variable in the model is the density of trees over some finite size region. This density  can take values from 0 to 1, and can be thought to represent the stress state of the plate. The random addition of trees can be considered, in the coarse grained sense, as a smooth increase of the stress in the system, caused by external tectonic loading. When stress is high in some portion of the system, it can be abruptly reduced by a random lightning to some small value, through the occurrence of a fire that is interpreted in this view as an EQ. The microscopic stochastic laws of the DS model prevent it for reaching a globally synchronized oscillatory state.
The wide distribution of fires in the DS model corresponds to a wide distribution of earthquakes. However the exponent $\tau\sim 1.2$ observed in the DS model does not  correspond to the exponent for EQs, around 1.7. Also, as the system lacks an internal dynamics, ASs are not produced. However, a physically motivated mechanism to produce ASs also produces a realistic GR law, as I will explain now.

\begin{figure}[h]
\includegraphics[width=8cm,clip=true]{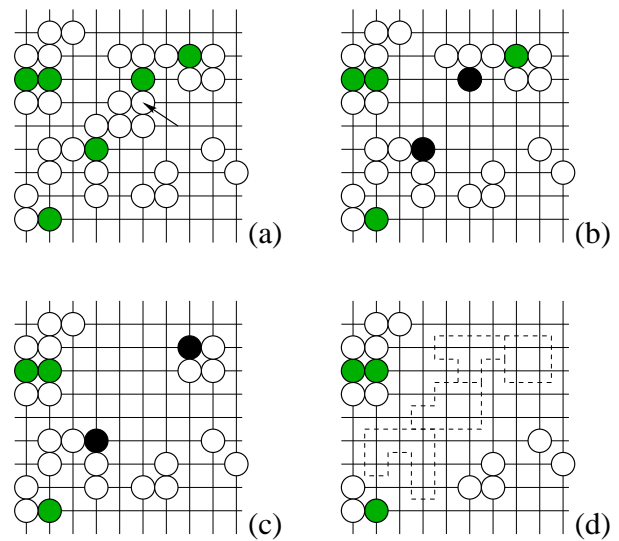}
\caption{(Color online) Propagation of fires in the modified DS model. White (gray) circles indicate A (B) trees. (a) In some initial configuration, a lightning strikes the site indicated by the arrow. (b) The state after the instantaneous burning of the connected cluster of A sites. The two B trees that are highlighted as black have become ``active". (c) One of the active B sites is chosen at random and propagates fire to its neighbor trees. Additional B sites may become active. (d) Final stage when there are no more active B sites. Along the whole process, fires of sizes 5, 4, 4, and 6 have occurred (contours are highlighted in (d)). The total size of the burnt cluster is thus 19.
}
\label{f1}
\end{figure}

I propose a modified DS model by introducing a second species of trees, called B trees (the original ones are noted A trees). The rules of the modified model are the following. In the insertion step,
each time a site is chosen for insertion and if this site is empty, it becomes occupied by an B tree with probability $w$, and by a A tree with probability 
$1-w$. The value of $w$ will be assumed to be small ($w\ll 1$) . In the burning step, A trees burn and propagate fire instantaneously to all neighbors, irrespective if they are A or B trees. However, B trees are supposed to propagate fire to neighbors only after some random time of the order of a new time scale $t_1$. This time scale $t_1$ is supposed to be much smaller than the insertion time $t_0$, yet  much larger than the propagation of fire through A trees. 
An example of the time sequence of fires in this modified DS model is shown in Fig. \ref{f1}. In the initial configuration, there is a given distribution of A and B trees partially filling the lattice, and a lightning strikes at some random site. This produces an instantaneous fire that burns all trees 
connected to the original site, and that ``activates" the B trees highlighted in (b), which retard the propagation for some time of the order of $t_1$. As these times for individual B trees are considered to be random, for the implementation I simply choose one of the active B trees randomly and start a secondary fire (c). The process is continued until no tree is burning (d). At this stage, the insertion process is continued. 

Time sequences of fires consist now of ``clusters" of events that are triggered by lightnings, and are separated by the time $t_0$. Each cluster is formed by events separated by  time intervals of the order of $t_1$. The precise values of these times will be important in a realistic consideration of ASs. Here I simply consider that these secondary fires are exactly triggered one every time $t_1$. 
In the EQ analogy, secondary fires are the ASs, and active B trees are their epicenters. The typical number of ASs observed depends on the parameter $w$, fixing the ratio between B and A trees at insertion. Clusters of events correspond to the initial shock and all its ASs.
The assumed condition $t_1/t_0\to 0$ allows to clearly identify ASs in the model, a situation that is not fulfilled in actual seismicity.

\begin{figure}[h]
\includegraphics[width=8cm,clip=true]{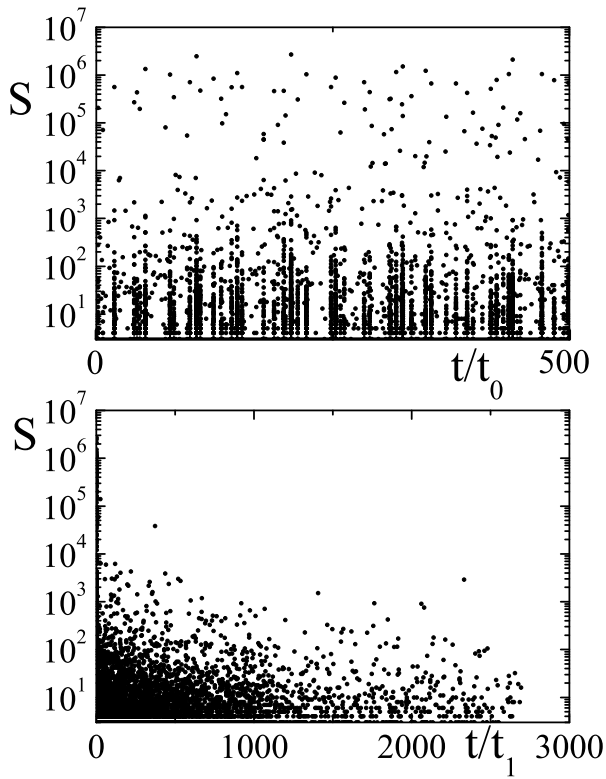}
\caption{(a) Temporal sequence of fires, in the modified DS model on a system of size 4000$\times$ 4000 system, with $r=5\times 10^4$ and $w=10^{-3}$. Events that occur at the same value of $t/t_0$ form what I have called a cluster, corresponding to the initial fire, generated by lightning, and all secondary fires propagated through B trees.
In (b), the same events are plotted as a function of its internal time within the cluster. This time is set to 0 for the first event of the cluster, and increases by $t_1$, at each  secondary fire.
}
\label{f2}
\end{figure}

\begin{figure}[h]
\includegraphics[width=8cm,clip=true]{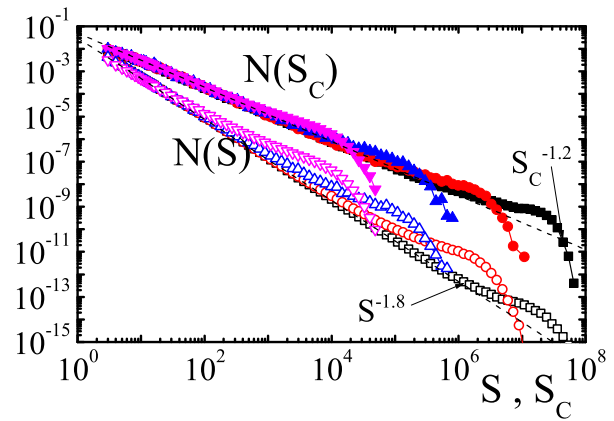}
\caption{(color online) Statistics of events for simulations with $r=10^ 3$, $10^ 4$, $10^ 5$, and $10^ 6$ (from left to right), and $w=0.01$ (system size is up to $20000\times 20000$). Lower curves correspond to individual events, whereas the upper ones are the results for the statistics of clusters. Statistics of clusters coincides with the result for single events in the case $w=1$.
}
\label{f3}
\end{figure}

A temporal sequence of events for the modified DS model is presented in Fig. \ref{f2}. The clustered structure of the sequence is apparent.  Secondary fires to a given initial lightning occur at the same value of $t/t_0$, and thus they appear on the same vertical line. The plot in panel (b) shows the same events, but plotted as a function of the internal time within the cluster $t/t_1$, with the value of $t$ set to 0 at the initial lightning of every cluster. 
Size distribution of events are presented in Fig. \ref{f3} with open symbols. The presence of B trees generates a size distribution with a $\tau$ exponent in the range 1.7-1.8, well different from the original slope ($\tau\simeq 1.2$) (more systematic results are presented in the supplementary material), and comparable to the value of the actual GR law. The slope change generated by the presence of B trees (by the existence of ASs, in the seismic perspective) is the main results of this work, and I will now explain its origin.

First of all, I notice that in addition to calculate the size distribution of individual events, it is also interesting to calculate the size distribution of clusters. Namely, I define the cluster size $S_C$ as the sum of sizes of the initial event and all its secondary fires. The size of the cluster $S_C$ is exactly what we had obtained as a single event if B trees also propagate fire instantaneously, or equivalently, if there are no B trees at all ($w=0$). In fact, the distribution of clusters, plotted in Fig. \ref{f3} with full symbols, coincides with the distribution in the original model\cite{grassberger}. Thus we see that the effect of B trees is to fragment the clusters in pieces that burn at different times. This fragmentation process, at the same time that generates secondary fires, produces the change in the exponent of the size distribution. 

It is useful to have an unambiguous characterization of this effect in a simpler model than the DS forest fire. In fact, in the DS model the spatial
distribution of trees is highly correlated, and rigorous results for the size distribution of avalanches are not available. However the analysis becomes much simpler if we consider the propagation of fires on a totally uncorrelated distribution of A and B trees, with spatial densities $\rho_A$ and $\rho_B$ (which now have to be fixed by hand). This problem can be studied with the tools of standard percolation theory.\cite{stauffer} If no B trees are present ($\rho_B=0$), events correspond to those that occur in a site percolation problem with probability $\rho_A$. As 
$\rho_A \to p_c\simeq 0.5927$, the events become power law distributed with an exponent $187/91-1\simeq 1.055$.\cite{stauffer} In the presence of B trees, we can calculate again
the distribution of single events and the distribution of clusters. Clusters are distributed as if B trees are not present, i.e, they correspond to normal percolation, with a cut off fixed by the total density $\rho_A+\rho_B$ ($<p_c$, to avoid an infinite fire). For single events instead, we have the same fragmentation effect discussed in the context of the DS
 model. The clusters become diluted with a fraction $\rho_B/(\rho_A+\rho_B)$ of B trees, generating a fragmentation effect. The size distribution of single events is the size distribution of these fragments. The effect of fragmenting a percolation cluster by the removal of a small fraction of sites was studied in \cite{gyure}, where it was shown that close to the percolation threshold, the removal of a single site generates fragments that are distributed according to a power law with an exponent $\phi$, the fragmentation exponent, that it was found to be $\phi\simeq 1.60$ in two dimensions. 

Results of numerical simulations in the two species version of the percolation problem (see results in the supplementary material) confirm this view: the inclusion of a small fraction of B trees changes the distribution of events to a new power law with a $\tau\simeq 1.60$. 
A perfect power law is obtained in the limit $\rho_A+\rho_B \to p_c$, $\rho_B/\left[p_c-(\rho_A+\rho_B)\right]\to const$. If $\rho_B$ is kept finite as $\rho_A+\rho_B \to p_c$, a non critical distribution is obtained, with maximum size of the fragments  controlled by the value of $\rho_A$.

The present analysis of the modified DS model and the percolation limit points out a simple mechanism that can justify the observed value of the $b$ exponent of the GR law in the context of a single planar fault situation. It suggests that the observation of this value is closely related to the presence of aftershocks.
It also points out a few possibilities to observe experimental signatures that would be a consequence of this mechanism. One of them is the fact that the statistics of clusters must display a less steep decay than that of individual EQs. 
An actual  verification of this fact would imply the unambiguous classification of events in clusters, which is a non trivial task, since  the $t_0$ and $t_1$ time scales are not well separated in actual seismicity. However, there have been promising attempts
to group the EQs in clusters according to appropriate metrics in the space-time-magnitude parameter space (see for instance \cite{baiesi}), that may serve to this purpose. A careful study of experimental data is needed to advance in this direction.

\begin{figure}[h]
\includegraphics[width=8cm,clip=true]{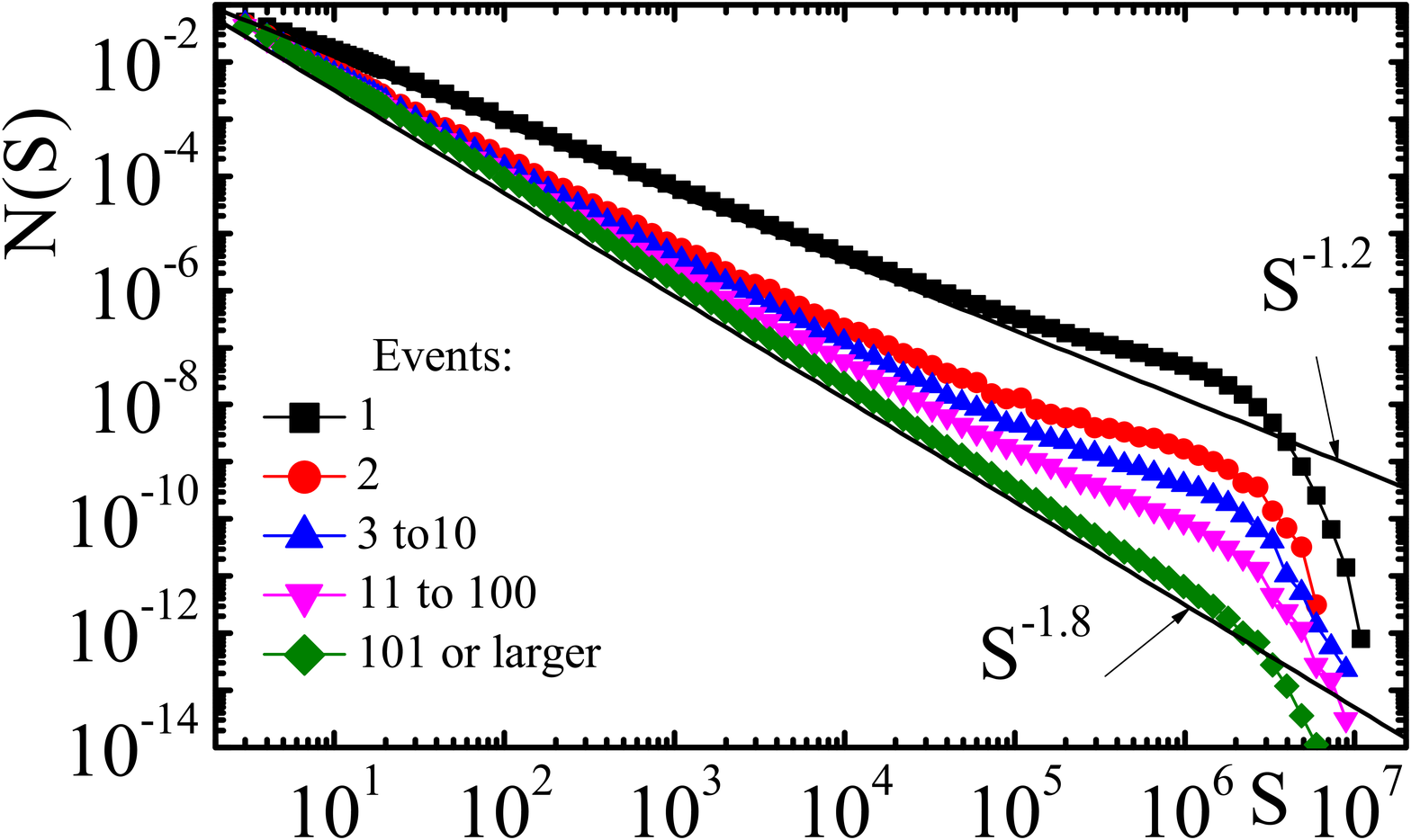}
\includegraphics[width=8cm,clip=true]{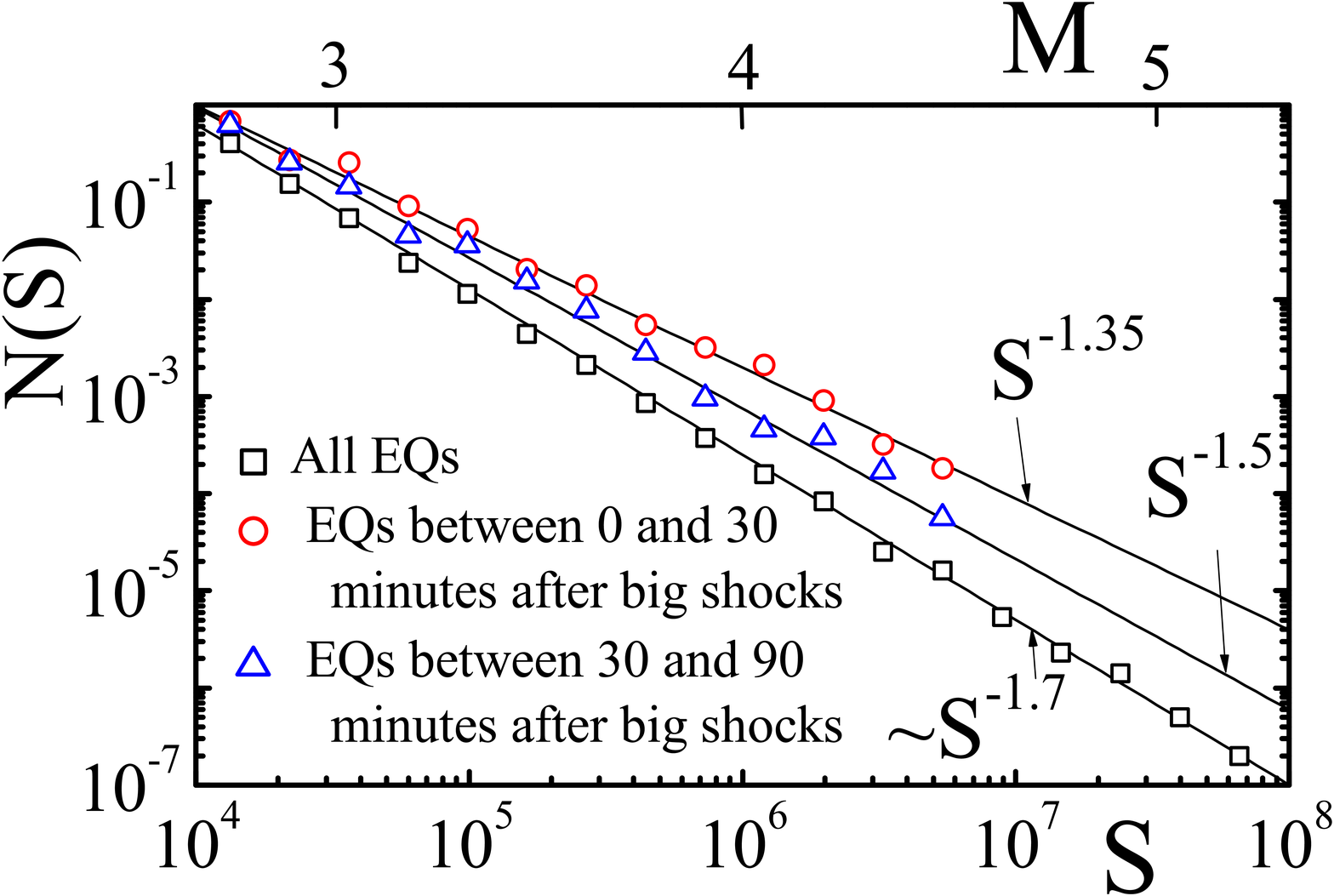}
\caption{(color online) (a) Partial size distribution of events in the modified DS model ($r=10^5$, $w=0.01$), restricted to the indicated events in the clusters. The distribution has an exponent close to that of the original model for the first event, and becomes progressively steeper as successive secondary fires are considered. (b) An actual example of this phenomenon (details are
provided in the supplementary information). The lower points display the distribution of EQs in Southern California during a 20 years time period. The upper curves correspond to the distributions of events that occurred in the two time windows indicated, after EQs of magnitude 4.5 or larger. The behavior is qualitatively similar to that in (a). Straight lines are for reference only.
}
\label{f4}
\end{figure}

A second signature indicating that the proposed mechanism may be relevant in actual seismicity is the following. If we take the events as a function of its internal time within an AS sequence (Fig. \ref{f2}(b)) and calculate the size distributions in small windows of the internal time, we obtain the results in Fig. \ref{f4}. It is seen that the decaying of the distribution has an exponent similar to that of the whole clusters for the first event, and becomes progressively steeper as successive events within the sequence are considered. This behavior can be understood by considering again the simpler percolation scenario. In that case, the very first event in each cluster is taken from a distribution with a density $\rho_A$ of occupied sites, and this has to follow a decaying law with the smaller $\tau$ ($\simeq 1.055$ in the percolation case). Successive ASs progressively drive the distribution towards the new, larger $\tau$ value.

This effect is suitable for experimental verification. 
In fact, EQ catalogs are dominated, after large magnitude events, by the ASs to those events. Although other events that are not ASs certainly occur, they are few in the first hours after large events. So they make a minor contribution that can be neglected in a first approach. In Fig. \ref{f4}(b) I show a preliminary analysis of the seismicity in Southern California over a 20 years period.  The tendency to display a 
weaker decay right after large events is clear from this figure, and gives support to the present proposed mechanism.

Summarizing, I have considered an analogy between a forest fire model and 
the stress in a single planar fault. The inclusion of a second tree species that delays propagation of fire, was shown to be analogous to include the possibility of ASs in the seismic counterpart.
In the modified model, EQs (or fires) appear in clusters formed by an initial event and all its ASs. The change of exponent of the GR law is a consequence of the fact that the clusters are fragmented by the existence of ASs. 
An approximate case that can be analyzed with standard percolation theory gives analytical support to this scenario. Also, observable signatures of this mechanism have been pointed out, and a preliminary analysis of actual seismic data where one of these signatures shows up has been presented.


I thank Alberto Rosso and Sebasti\'an Risau for helpful discussions.
This research was financially supported by Consejo Nacional de Investigaciones Cient\'{\i}ficas y T\'ecnicas (CONICET), Argentina. Partial support from
grant PIP/112-2009-0100051 (Argentina) is also acknowledged.

\vspace{4cm}

\begin{widetext}

{\center{\bf Supplemental material to: A forest-fire analogy to explain the {\em b}-value of the Gutenberg-Richter law for earthquakes}}\\

{\bf A: Size distribution of events for an uncorrelated distribution of A and B trees}\\

Let us assume a random distribution of A and B trees on a square lattice, with densities $\rho_A$ and $\rho_B$.
An equivalent, more convenient set of variables is: $p\equiv\rho_A+\rho_B$, and $\varepsilon\equiv\rho_B / \rho_A$ (note that $\varepsilon$ is related to $w$ introduced in the main paper by $w=\varepsilon/(1+\varepsilon)$).
We want to calculate $N(S,p,\varepsilon)$, the probability distribution of fires  as a function of their size $S$, for given values of $p$ and $\varepsilon$.  
This function displays scaling properties as $p\to p_c\simeq 0.5927$ and for small $\varepsilon$. An appropriate scaling form is given by
\begin{equation}
N(S,p,\varepsilon)=N^0\left(S(p_c-p)^{1/\sigma},\varepsilon/(p_c-p)\right),
\label{eq1}
\end{equation}
where a parameter dependent normalization factor has not been included. The value of the exponent $\sigma$ is exactly known in 2D percolation, and is given by $\sigma=36/91$.
Numerical results adjust very well to this form. They are shown in Fig. S\ref{f5}.

The numerical simulation in the present percolation case, is made in the same way as described in Fig. 1 for the modified DS model, but using an uncorrelated distribution of A and B trees.
In order to to speed up the simulations, only the trees that are burnt are actually generated. 
This is done by starting a cluster at the origin, and sequentially extending it, introducing A and B trees with their corresponding probabilities 
$\rho_A$ and $\rho_B$.

The distribution obtained for $\rho_B=0$ corresponds to the usual site percolation case. 
In particular, $N^0(x,0)\sim x^{-96/91}$ for small $x$, so for $p\to p_c$ we have
\begin{equation}
N(S,p,0)=N^0(S(p_c-p)^\sigma,0)\sim S^{-96/91}
\end{equation}
As $\rho_B$ becomes different from zero, the size distribution of events departs from the result of standard percolation and starts to develop a region at low events sizes where a distribution according to the fragmentation exponent $\phi\simeq 1.60$ is displayed, i.e, 
for constant $\varepsilon$, and for $p\to p_c$ we get 
\begin{equation}
N(S,p,\varepsilon)\sim S^{-\phi}
\end{equation}
and this behavior remains correct for $S\lesssim (p_c-p)^{-\sigma}$. This means that 
a strict  power law with the fragmentation exponent is obtained in the double limit $p\to p_c$, $\varepsilon/(p_c-p) \to {\mbox const.} \ne 0$, i.e., the density of B trees has to be reduced to zero at the same time that the total density goes to $p_c$. If the limit $p\to p_c$ is taken keeping a finite value of $\rho_B$ the limiting distribution has an exponential cutoff at large events sizes that is controlled by the value of $\rho_A<p_ c$.\\

\begin{figure}[h]
\includegraphics[width=14cm,clip=true]{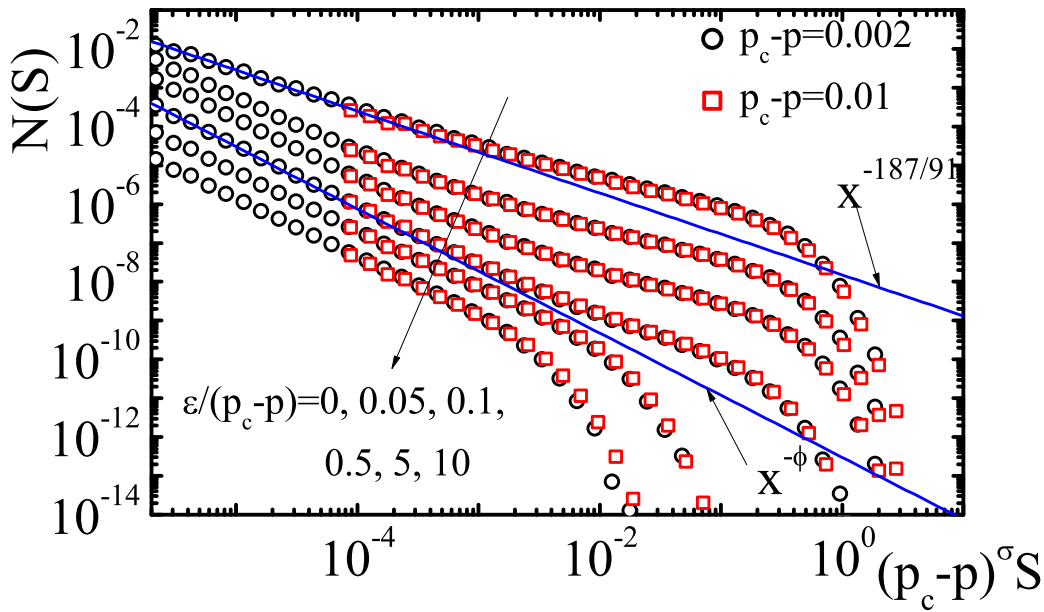}
\caption{Results of simulations in the percolation case, at two different values of total density $p\equiv \rho_A+\rho_B$, and for different relative density of A and B trees $\varepsilon=\rho_B / \rho_A$. The results follow the scaling form presented in Eq. (\ref{eq1}) (with $\sigma=36/91$), which allows to conclude that a pure power law with an exponent $\phi\simeq 1.60$ is obtained in the double limit $p\to p_c$, $\varepsilon/(p_c-p) \to {\mbox const.} \ne 0$
}
\label{f5}
\end{figure}

{\bf B: Simulations of the modified DS model}\\

Here I present additional results for the modified DS model. They are shown in Fig. S\ref{f6}, and correspond to simulations with progressively larger values of $r$, and different values of the probability of B trees at insertiuon $w$. For non-zero $w$, the main dependence of $N$ with $S$ is absorbed by plotting $NS^\tau$ 
and using $\tau=1.8$. A tentative scaling is made by plotting  the horizontal axis as $S/r^\lambda$, using $\lambda=1.08$ which is the expected
scaling in the original DS model. We see a non-perfect scaling of the results, with systematic deviations that remain even for the largest values of $r$ used, as it was the case of the original DS model \cite{grassberger,jensen2}.

For $w$ in the approximate range $10^{-3}-10^{-2}$ there is a progressively wider region when $r$ is increased in which $NS^{1.8}$ is practically constant, meaning that a distribution with the modified exponent $N(S)\sim S^{-1.8}$ is obtained.
By comparing Figs. S\ref{f5} and S\ref{f6} we can conclude that,
in addition to the lack of true criticallity, a further difference between the results for the modified DS model, and those in the percolation limit is that the more accurate scaling of the data in Fig. S\ref{f6} is obtained by keeping a fixed value of $w$ as $r$ is increased, contrary to the percolation case, in which the density of B trees has to be reduced to zero at the same time that $p\to p_c$.
The origin of this difference (and whether it persists for larger values of $r$ or not) deserves further investigation, but it can be anticipated that the answer is contained in the structure of the fires in the DS model compared with the normal percolation clusters.
More systematic simulations in larger systems would be helpful to adress this issue, but I note that 
 the distribution of events following $N(S)\simeq S^{-1.8}$ extends already through a range of $S$ values that is comparable to the experimental range for EQs.\\

\begin{figure}[h]
\includegraphics[width=12cm,clip=true]{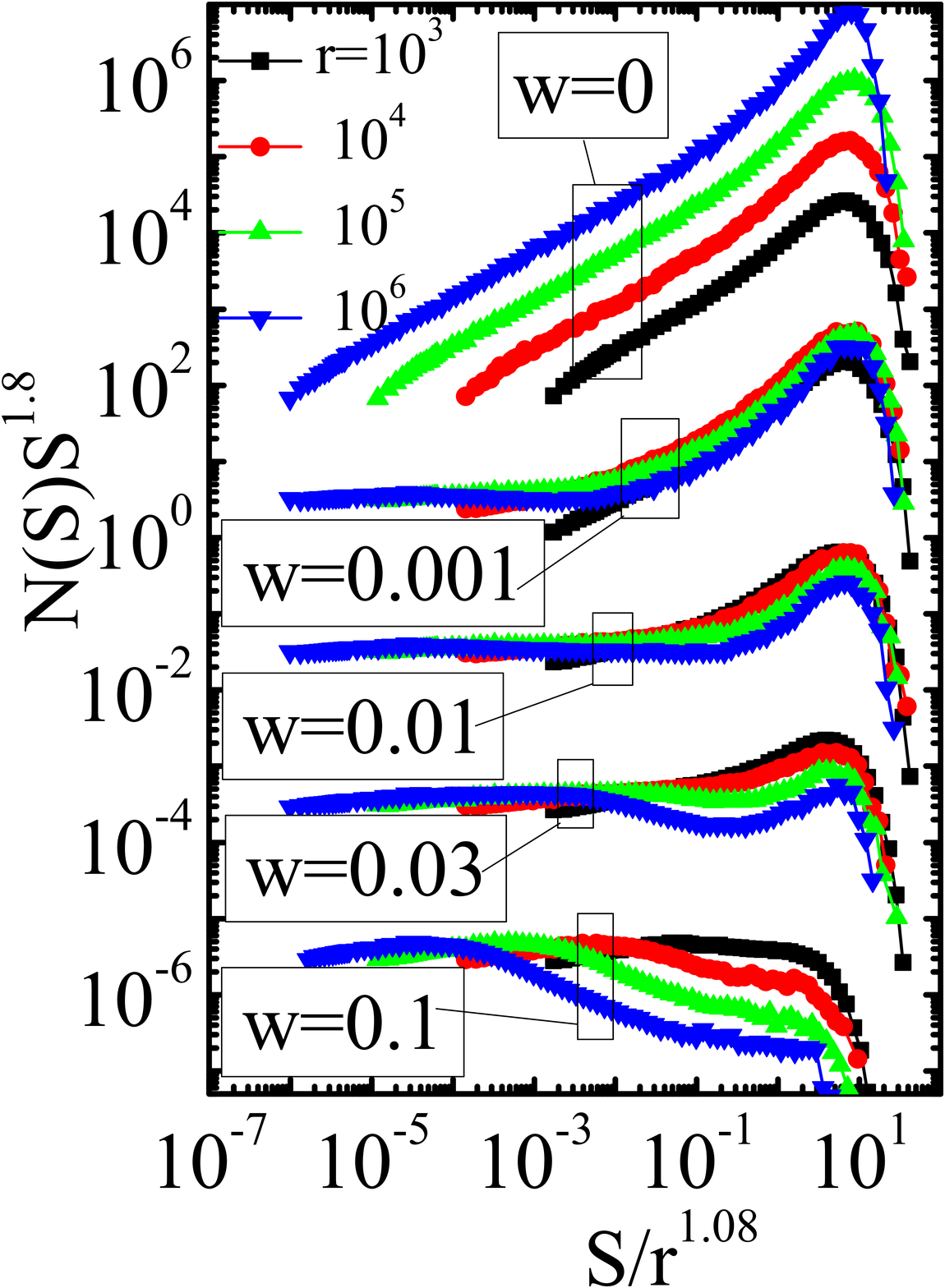}
\caption{Results of simulations of the modified DS model as a function of $r$, and for different values of $w$. Groups of curves at a same value of $w$ have been displaced vertically by the same factor, for clarity.
}
\label{f6}
\end{figure}

{\bf C: Details on the data in Fig. 4(b)}\\

Here I give details on the data analysis to produce Fig. 4(b). The starting point is the list of all EQs (charaterized by magnitude and time)  in the region of Southern California (between 115$^\circ$ - 119$^\circ$ W, and 32$^\circ$ - 37$^\circ$ N), that occurred between 1/1/1980 and 5/20/2008 with magnitude larger that 2.5 (according to the ANSS database, http://earthquake.usgs.gov/monitoring/anss/, about 26000 events are counted). The number of EQs in small magnitude windows were calculated and the results are shown by the squares. Next, the 108 events with magnitude larger that 4.5 were considered, and the statistics was limited to the EQs with time of ocurrence between $t_A$ and $t_B$ after any of these 108 large EQs. The results are plotted as circles ($t_A=0$, $t_B=30$ min, 970 events are counted) and triangles ($t_A=30$ min, $t_B=90$ min, 930 events are counted). Note that these two partial distribution are limited to magnitudes lower than 4.5, since larger events are counted as main shocks.

\end{widetext}


\begin{thebibliography}{4}

\bibitem{soc}P. Bak, C. Tang, K. Wiesenfeld, Phys. Rev. Lett. {\bf 59}, 381 (1987), P. Bak, {\it How Nature Works} (Oxford University Press, Oxford,
1997); H. J. Jensen, {\it Self-Organized Criticality} (Cambridge University Press, Cambridge, 1998).

\bibitem{gr}B. Gutenberg and C. F. Richter, Ann. Geophys. (C.N.R.S) {\bf 9}, 1 (1956).

\bibitem{scholz}C. H. Scholz, {\em The Mechanics of Earthquakes and Faulting},  (Cambridge University Press, Cambridge, England, 2002).

\bibitem{kawamura}H. Kawamura, T. Hatano, N. Kato, S. Biswas, and B. K. Chakrabarti,
Rev. Mod. Phys. {\bf 84}, 839 (2012).

\bibitem{bk} R. Burridge and L. Knopoff, Bull. Seismol. Soc. Am. {\bf 57}, 341 (1967).

\bibitem{ofc} A. Olami, H. J. S. Feder, and Christensen, Phys. Rev. Lett. {\bf 68}, 1244 (1992).

\bibitem{hainzl}S. Hainzl, G. Zoller, and J. Kurths, J. Geophys. Res. {\bf 104}, 7243 (1999);
S. Hainzl, G. Zoller, and J. Kurths, Nonlin.  Proc. Geophys. {\bf 7}, 21 (2000).


\bibitem{jagla_jgr}E. A. Jagla and A. B. Kolton, J. Geophys. Res.,   {\bf 115} B05312 (2010).

\bibitem{jagla_pre}E. A. Jagla, Phys. Rev. E {\bf 81}, 046117 (2010).




\bibitem{ds} B. Dr\"ossel and F. Schwabl, Phys. Rev. Lett. {\bf 69}, 1629 (1992).

\bibitem{pacheco}A. Tejedor, J. B. G\'omez, A. F. Pacheco, Phys. Rev. E {\bf 79}, 046102 (2009).

\bibitem{jensen} P. Sinha-Ray and H. J. Jensen, Phys. Rev. E {\bf 62}, 3215 (2000).

\bibitem{turcotte0} W. I. Newman and D. L. Turcotte, Non. Proc. Geophys. {\bf 9}, 453 (2002).

\bibitem{grassberger}P. Grassberger, New J. Phys. {\bf 4}, 17 (2002).

\bibitem{jensen2}G. Pruessner  and H. J. Jensen, Phys. Rev. E   {\bf 65}, 056707 (2002)

\bibitem{stauffer}D. Stafuffer and A. Aharony, {\em Introduction to Percolation Theory},  (Taylor and Francis, London, 1991).

\bibitem{gyure}M. F. Gyure and B. F. Edwards, Phys. Rev. Lett. {\bf 68}, 2692 (1992).

\bibitem{baiesi} M. Baiesi and M. Paczuski, Phys. Rev. E {\bf 69}, 066106 (2004).

\end{thebibliography}
\end{document}